\documentclass[preprint,aps,nofootinbib]{revtex4-1}

\usepackage{array}
\usepackage{booktabs}
\usepackage{tabu}
\usepackage{footnote}
\usepackage{dcolumn}
\usepackage{amsmath}
\usepackage{amsfonts}
\usepackage{amssymb}
\usepackage{graphicx}
\usepackage{subfigure}
\usepackage{graphicx}
\usepackage{dcolumn}
\usepackage{bm}
\usepackage{xcolor}

\def\be{\begin{equation}}
  \def\ee{\end{equation}}
\def\bea{\begin{eqnarray}}
\def\eea{\end{eqnarray}}
\def\f{\frac}
\def\n{\nonumber}
\def\l{\label}
\def\p{\phi}
\def\o{\over}
\def\R{\rho}
\def\pa{\partial}
\def\om{\omega}
\def\na{\nabla}
\def\P{\Phi}

\begin{document}

\title{Constant-roll brane inflation}

\author{Abolhassan Mohammadi$^a$}
\email{a.mohammadi@uok.ac.ir;abolhassanm@gmail.com}
  \author{Tayeb Golanbari$^a$}
  \email{t.golanbari@uok.ac.ir; t.golanbari@gmail.com}
   \author{Salah Nasri$^{b, c}$}
   \email{snasri@uaeu.ac.ae}
           \author{Khaled Saaidi$^a$}
        \email{ksaaidi@uok.ac.ir}
  \affiliation{
$^a$Department of Physics, Faculty of Science, University of Kurdistan,  Sanandaj, Iran.\\
$^b$Department of physics, United Arab Emirates University, Al-Ain, UAE.\\
$^c$ The Abdus Salam International Centre for Theoretical Physics, Strada
Costiera 11, I-34014, Trieste, Italy.
}

\date{\today}

\def\be{\begin{equation}}
  \def\ee{\end{equation}}
\def\bea{\begin{eqnarray}}
\def\eea{\end{eqnarray}}
\def\f{\frac}
\def\n{\nonumber}
\def\l{\label}
\def\p{\phi}
\def\o{\over}
\def\R{\rho}
\def\pa{\partial}
\def\om{\omega}
\def\na{\nabla}
\def\P{\Phi}

\begin{abstract}
The scenario of constant-roll inflation in the frame of the RSII brane gravity model is considered. Based on the scenario, the smallness of the second slow-roll parameter is released and it is assumed as a constant which could be of the order of unity. Applying the Hamilton-Jacobi formalism, the constancy of the parameter gives a differential equation for the Hubble parameter which leads to an exact solution for the model. Reconsidering the perturbation equations we show  there are some modified terms appearing in the amplitude of the scalar perturbations and in turn in the scalar spectral index and tensor-to-scalar ratio. Comparing the theoretical results of the model with observational data, the free parameters of the model are determined. Then, the consistency of the model with the swampland criteria is investigated for the obtained values of the free parameters. As the final step, the attractor behavior of the model is considered.

\end{abstract}
\pacs{............}
\maketitle

\section{Introduction}
Inflation is an exponential expansion of space in the very early universe during an extremely short period of time. This scenario has received observational support \cite{Planck:2013jfk,Ade:2015lrj,Array:2015xqh,Akrami:2019izv,Akrami:2018odb} which makes it one of  the cornerstones of physical  cosmology, rendering  any model  for the evolution of the universe  incomplete without  the inflation phase. The first realistic  inflationary scenario   was proposed  four decades ago   as  a solution to two  of the shortcomings of the  hot big bang model\cite{Guth:1980zm,Linde:2000kn,Linde:2005ht,Linde:2005vy,Linde:2004kg}, namely the flatness and the horizon problems, and  since then became   the  leading  paradigm for the early universe.

So far, many inflationary models have been introduced based on the slow-roll assumptions where the inflaton, a scalar field,  slowly rolls down its  potential.  It is described  by  two dimensionless parameters, known as the slow-roll parameters, which their smallness  during inflation  guarantees an almost flat   potential  \cite{Riotto:2002yw,Baumann:2009ds,Weinberg:2008zzc,Lyth:2009zz}. Example of such Models include non-canonical inflation \cite{Barenboim:2007ii,Franche:2010yj,Unnikrishnan:2012zu,Gwyn:2012ey,Rezazadeh:2014fwa,Cespedes:2015jga,Stein:2016jja,
Pinhero:2017lni,Teimoori:2017wbx}, tachyon inflation \cite{Fairbairn:2002yp,Mukohyama:2002cn,Feinstein:2002aj,Padmanabhan:2002cp}, DBI inflation \cite{Spalinski:2007dv,Bessada:2009pe,Weller:2011ey,Nazavari:2016yaa,Amani:2018ueu}, G-inflation \cite{maeda2013stability,abolhasani2014primordial,alexander2015dynamics,tirandari2017anisotropic}, and warm inflation \cite{berera1995warm,berera2000warm,taylor2000perturbation,hall2004scalar,BasteroGil:2004tg,Sayar:2017pam,Akhtari:2017mxc,
Sheikhahmadi:2019gzs}. However,  a different inflationary scenario has been proposed  very recently which goes beyond   the slow-roll approximation, where  the second slow-roll parameter does not have to be smaller than unity and can be  a constant\cite{Kinney:2005vj, Namjoo:2012aa}. This  constant-roll inflation scenario (CRI),   attracted a lot of interest among cosmologists  as an alternative  way for the  inflation phase to take place \cite{Anguelova:2017djf,Gao:2018tdb,Odintsov:2017hbk,Oikonomou:2017bjx,Karam:2017rpw,Nojiri:2017qvx,Motohashi:2017vdc,
Awad:2017ign,Odintsov:2017yud,Odintsov:2017qpp,Oikonomou:2017xik,Mohammadi:2018oku,Mohammadi:2019dpu,
Mohammadi:2018zkf,Motohashi:2019tyj,Odintsov:2019ahz,Mohammadi:2019qeu,Antoniadis:2020dfq,Cicciarella:2017nls}.

Inspired by string theory,  one can consider our observable universe to be  a $(3 +1)$  four-dimensional  hypersurface (the brane) embedded in a higher-dimensional spacetime (the bulk). We consider a five-dimensional space and  assume that all standard model particles  to be confined to the brane, and only gravity is allowed to  propagate in  the fifth dimension. The most popular are the ADD and the RS models   \cite{ArkaniHamed:1998rs, Randall:1999ee} which  were proposed in an attempt to solve the hierarchy problem between the Planck  scale  and  the electro-weak scale.
 Furthermore, in some inflationary models in the context of the brane world scenario, the inflaton potential  arises  naturally from higher-dimensional gravity \cite{Binetruy:1999hy,Binetruy:1999ut,Brax:2004xh,Maartens:2010ar,Saaidi:2010ju,Saaidi:2012ri,
Davis:2005au,Saaidi:2012rj,Visinelli:2017bny,Vagnozzi:2019apd,Fichet:2019owx,Pardo:2018ipy,
Chakravarti:2019aup,Banerjee:2019nnj,Mack:2019bps,Garcia-Aspeitia:2020snv}  and yields  interesting cosmological implications\cite{maartens2000chaotic,golanbari2014brane,Banerjee:2017lxi,Elizalde:2018rmz,Paul:2018jpq}. 

There has been a growing interest in applying the CRI scenario to many inflationary models which, depending on the details of the model, results in some modification of the Friedmann equation. In the brane world scenario, the Friedmann equation will contain both quadratic and linear terms,   which in the high energy regime (i.e. $\rho \gg \lambda$) the linear term can be ignored. In this case, unlike the standard  four-dimensional cosmology, the Hubble parameter behaves as $H \propto \rho$ rather than $H \propto \sqrt{\rho}$, a novel aspect of the CRI scenario in this context. It is expected that this modification affects relevant parameters and observable of the inflation phase,  including the slow-roll parameters, the shape of the potential, end time of inflation,  and the magnitude of the inflaton. Due to this novel feature of the Friedmann equation, it is important to consider the CRI scenario in the framework of the brane world and study the new features that may arise in this case. \\
Another motivation for considering the brane world  comes from the swampland  conjectures \cite{Obied:2018sgi,Ooguri:2018wrx,Garg:2018reu}, which can be used as criteria  to distinguish effective field theories (EFT) that can be  UV-completed to a quantum theory of gravity. The first criterion requires the field range  traversed by the fields to  be bounded from above by a value of  order one, whereas the second criterion imposes
a lower bound on the gradient of the potential. The latter bound is in direct tension with  inflation where the first slow-roll parameter $\epsilon = M_p^2 V'^2 / 2 V^2$ must be smaller than one. Thus,  some inflationary models are  not compatible  with these criteria, and hence can not be embedded into a  consistent theory of quantum gravity.  However, inflationary models in the brane-world scenario have the potential to evade the swampland constraints\cite{Kehagias:2018uem,Kinney:2018kew,Matsui:2018bsy,Lin:2018rnx,Dimopoulos:2018upl,Das:2018rpg,
Kinney:2018nny,Geng:2019phi}, and hence it will be interesting to investigate  inflation in  this framework and its implications.
\\
The paper is organized as follows.  In Sec II, the main evolution equations of the model are given. The scenario of the constant-roll inflation is discussed in the frame of brane world in Sec.IV.  The exact solutions for the model are obtained in Sec.III and the main dynamical parameters are obtained in terms of the scalar field. In Sec.IV, the cosmological density  perturbations are considered, and the consistency of the model with the observational data and swampland criteria are investigated in Sec.V and VI, respectively. As the last step, the attractor behavior of the solution is studied in Sec.VII. The results are summarized in Sec.VIII.

\section{The model}
The action for the brane world is given by
\begin{equation}\label{action}
  S = \int d^5x \sqrt{-g} \left( {M_5^3 \over 2} \; \mathcal{R} + \Lambda_5 \right) \; +
       \int d^4x \sqrt{-h} \left( \; \mathcal{L} + \lambda \right)
\end{equation}
where the  first integral  represents  the action of the bulk and the second one corresponds to the brane, $\mathcal{R}$ is the Ricci scalar related to the five-dimensional metric $g_{AB}$, $g$ and $h$  denote  the determinants of the  metric  on the five-dimensional space and the brane, respectively, $\Lambda_5$ the five-dimensional cosmological constant, $\mathcal{L}$ the lagrangian of the matter fields, and $\lambda$ the brane tension. \\
Taking variation of the action with respect to the metric yields  the field equation
\begin{equation}\label{field equation}
 G_{\mu\nu} = -\Lambda_4 g_{\mu\nu} + \left({8\pi\over M_{4}^2}\right) T_{\mu\nu}  + \left({8\pi\over
M_5^3}\right)^2 \Pi_{\mu\nu} - E_{\mu\nu}\ ,
\end{equation}
Here   $T_{\mu\nu}$  is  the energy-momentum tensor of the matter on the brane, $\Pi_{\mu\nu}$ a tensor that  includes the terms quadratic in $T_{\mu\nu}$, and $E_{\mu\nu}$  represents the projection of Weyl tensor on the brane  which portray the effects of the bulk graviton on the dynamical evolution of the brane. Assuming  the geometry of the universe  to be  described by a  five-dimensional Friedmann–Lemaitre–Robertson–Walker (FLRW) metric \begin{eqnarray} \label{FLRW}
ds^2_5 = -dt^2 + a^2 \delta_{ij} dx^i dx^j + dy^2  ,
\end{eqnarray}
the Friedmann equation reads
\begin{equation}
H^2={\Lambda_4 \over 3} + \left( {8\pi \over 3M_4^2} \right) \; \rho + \left({4\pi\over 3M_5^3}\right)^2 \; \rho^2  +{\mathcal{C} \over a^4} .
\end{equation}
with  $\Lambda_4$ is the cosmological constant of the brane, and $\mathcal{C}/a^4$ is known as the dark radiation\footnote{This is because of its  dependence on the scale factor  is the same  as the energy density of  radiation.}. The five and four-dimensional Planck masses in the above equation  are related as $M_4 = \sqrt{3 \over 4\pi \lambda} \; M_5^3$. \\

During inflation,  the dark radiation term gets  diluted, and  hence can be  neglected. Also,  here the RS fine-tuning is being used  to  set  the four-dimensional cosmological constant  to zero. Thus, the  Friedmann equation gets  reduced to
\begin{equation}\label{Friedmann}
H^2 = {8\pi \over 3M_4^2} \rho \left( 1+ {\rho \over 2\lambda} \right) ,
\end{equation}

Since all the matter fields  are confined on the brane, the  conservation  of energy in this expanding universe   is the same as in   standard cosmology, i.e.
\begin{equation}\label{conservation}
  \dot{\rho} + 3 H (\rho+p)=0.
\end{equation}
Using the above equation  and taking the time derivative of Eq.\eqref{Friedmann},  we obtain the second Friedmann equation
\begin{equation}\label{SFriedmann}
  \dot{H} = {-4 \pi \over M_4^2} \; \left( 1 + {\rho \over \lambda} \right) \; (\rho + p).
\end{equation}

Inflation is driven  the inflaton,  a scalar field $\phi$, that  is  confined  on the brane  and with   energy density and pressure
\begin{equation}
\rho = {\dot{\phi}^2 \over 2 } + V(\phi), \qquad p= {\dot{\phi}^2 \over 2 } - V(\phi)
\end{equation}
and  obeys  the  equation of motion 
\begin{equation}\label{EoM}
\ddot{\phi} + 3H \dot{\phi} + V'(\phi) = 0.
\end{equation}

It is widely common to consider the inflation at  the energy scale where the energy density is larger than the tension of the brane, i.e. $\rho \gg \lambda$. Therefore, the above Friedmann equations are reduced to
\begin{equation}
H^2 = \left( 4\pi \over 3M_5^3 \right)^2 \rho^2, \qquad \dot{H} = -3 \; \left( 4\pi \over 3M_5^3 \right) H \dot{\phi}^2
\end{equation}

\section{Constant-roll inflation}
In slow-roll inflationary models,  the inflaton rolls down its  potential   very slow  which can be  described in terms of  dimensionless  parameters
\begin{equation}
\epsilon = {-\dot{H} \over H^2}  ~~\text{and}~~~ \qquad \eta = {-\ddot{\phi} \over H \dot{\phi}}
\end{equation}
which satisfy the  conditions   $\epsilon < 1$ and $\eta < 1$,   known as slow-roll parameters  (SRP)\cite{Weinberg:2008zzc}.
Another  scenario is  the  constant-roll inflation  where the second slow-roll parameter is assumed to be   constant and can  be of order of unity:
\begin{equation}
\eta = {-\ddot{\phi} \over H \dot{\phi}} = \beta = \text{constant}
\end{equation}
The fact that $\eta$ is a constant results in a differential equation for the Hubble parameter that admits an exact solution for the model. For that,  we first obtain the time derivative of the scalar field from the second Friedmann equation by taking the Hubble parameter as a function of the scalar field, i.e. $H:=H(\phi)$, and write
\begin{equation}\label{phi-dot}
\dot{H} = \dot{\phi} H' ~~~ \qquad \Rightarrow \qquad
\dot{\phi} = {-1 \over 3} \left( 3M_5^3 \over 4\pi \right) {H' \over H}
\end{equation}
Then,  it follows the following differential equation for the Hubble parameter 
\begin{equation}
H H'' - H'^2 - \tilde{\beta} H^3 = 0, \qquad \tilde{\beta}= {4\pi \over 3M_5^3} \; \beta.
\end{equation}
and has   a  solution  given by
\begin{equation}\label{Hubble}
H(\phi) = {-\alpha \over 2\tilde{\beta}} \; \left[ 1 - \tanh\left( {\sqrt{\alpha} \over 2} \; (\phi + \phi_0) \right) \right]
\end{equation}
where  $\alpha$ and $\phi_0$ are  constants  of  integration. Note that since the Hubble parameter is positive and  the term $\tanh$ is smaller than one, the constant $\alpha$ must be negative.\\

Now that we have  the  expression  of  $H (\phi)$,  we  can derive $\dot{\phi} (\phi)$  and $V (\phi)$, and we get 
\begin{eqnarray}
\dot{\phi} & = & {M_5^3 \sqrt{\alpha} \over 4\pi} \; \tanh\left( {\sqrt{\alpha} \over 2} \; (\phi + \phi_0) \right) \label{phidot}\\
V(\phi) & = & \left( {M_5^3 \over 4\pi} \right)^2 \; {\alpha \over 2} \left[ {-3 \over \beta} + \left( {3 \over \beta} - 1 \right) \tanh^2\left( {\sqrt{\alpha} \over 2} \; (\phi + \phi_0) \right) \right]. \label{pot}
\end{eqnarray}
By integrating the equation of $\dot{\phi}$ above,    gives   the time evolution of  scalar field  as
\begin{equation}\label{phit}
\phi(t)+\phi_0 = {2 \over \sqrt{\alpha}} \; \sinh\left[ \exp\left( {M_5^3 \alpha \over 8\pi} \; (t+t_0) \right) \right]
\end{equation}

\subsection{Scalar field at the  horizon crossing time}
The  inflationary phase  will come to an  end when the  SRP $\epsilon (\phi)$  becomes equal to unity, i.e.
\begin{equation}\label{epsilon}
\epsilon(\phi_e) := 2\beta \; {\tanh^2\left( {\sqrt{\alpha} \over 2} \; (\phi_e + \phi_0) \right) \over 1-\tanh^2\left( {\sqrt{\alpha} \over 2} \; (\phi_e + \phi_0) \right)} = 1
\end{equation}
where $\phi_e$ is the value of the field at the exit of inflation, which can be determined by solving the above algebraic equation. With this we can quantify the amount of inflation the universe underwent, corresponding to
the number of e-fold from the beginning of inflation, the instant $t_i$, to the exit time $t_e$, and is  given by
\begin{equation}
N = \int^{t_e}_{t_i} H dt = \int^{\phi (t_e) \equiv \phi_e}_{\phi (t_i)\equiv\phi_i} {H \over \dot{\phi}} \; d\phi =
- {4\pi \over M_5^3} \; \int^{\phi_e}_{\phi_i} {H^2 \over H'} \; d\phi
\end{equation}
Substituting the solution we have obtained for the Hubble parameter, and after some manipulation we obtain
\begin{equation}
N = {-4\pi \over M_5^3 \tilde{\beta}} \; \ln\left( \tanh\left[ {\sqrt{\alpha} \over 2}\; (\phi+\phi_0) \right] \right)\Big|^{\phi_e}_{\phi_i} =
{2\pi \over M_5^3 \tilde{\beta}} \; \ln\left(
{\tanh^2\left[ {\sqrt{\alpha} \over 2}\; (\phi_i +\phi_0)\right] \over
\tanh^2\left[ {\sqrt{\alpha} \over 2}\; (\phi_e +\phi_0) \right] }
\right)
\end{equation}
or, equivalently
\begin{equation}\label{phistar}
\tanh^2\left[ {\sqrt{\alpha} \over 2}\; (\phi_i +\phi_0)\right] = {e^{2\beta N} \over 1-2\beta}
\end{equation}

\section{Cosmological perturbations}
In this section we consider the impact on the quantum perturbations as one of the most important predictions of  inflation which represents the main test that we have for verifying any inflationary model. The perturbations are usually divided into three types: scalar, vector, and tensor. Vector perturbations are usually ignored as  they depend on  the inverse of the scale factor and get diluted rapidly during inflation. Scalar perturbations are the seed for large scale structure formation  in  the universe. The tensor perturbations describe  the primordial gravitational waves   which  have not been detected yet and at present we have only   an upper bound on the tensor-to-scalar ratio. \\

The study of the cosmological perturbation in constant-roll inflation is a little different than in the slow-roll scenario. Since the second SRP, $\eta$,  might be of order unity, in calculating the scalar and tensor perturbations the terms $\eta^2$, $\epsilon \eta$, and $\epsilon \eta^2$ can  not be ignored. In this regard, the whole perturbation equations involving the second SRP should be reconsidered. In the following subsections, we are going to reconsider both scalar and tensor perturbations for any possible modification. \\

\subsection{Scalar perturbations}
To derive the perturbation parameters  we usually need to obtain the Mukhanov-Sasaki equation \cite{Baumann:2009ds,Weinberg:2008zzc,Sasaki:1983kd,Kodama:1985bj,Sasaki:1986hm,Mukhanov:1985rz,Mukhanov:1988jd,Lyth:2009zz}. For this matter, the action is computed up to the second order of the perturbation parameter. Following \cite{Baumann:2009ds,baumann2015physics}, the spatially flat gauge is used in which,  up to the leading order of $\epsilon$, the fluctuations in the geometry of the action could be ignored. Since the scalar field lives on the brane, we have the same perturbation equation as we have in the standard four-dimensional cosmology, that is
\begin{equation}\label{vDE}
v''_k(\tau) + \left( k^2 - {z'' \over z} \right) \; v_k(\tau) = 0
\end{equation}
where again $z$ has the same definition as $z^2 = a^2 \dot\phi^2 / H^2$. Therefore, after some algebraic  manipulations, the term $z''/z$  in the above equation can be expressed   as
\begin{equation}
{z'' \over z} = {1 \over \tau^2} \; \left( 2 + 6 \; \epsilon - 3 \beta - 9 \epsilon \; \beta + \beta^2 + 2 \epsilon \; \beta^2 \right).
\end{equation}
Making  the change of variables  $x=-k\tau$ and $f_k = v_k/{\sqrt{-\tau}}$,  Eq.\eqref{vDE} becomes    a Bessel differential equation as
\begin{equation}\label{Bessel}
{d^2 f_k \over dx^2} + {1 \over x} \; {df_k \over dx} + \left( 1 - {\nu^2 \over x^2} \right) \; f_k = 0,
\end{equation}
where we have used
\begin{equation}\label{nu}
{z'' \over z} = {\nu^2 - {1 \over 4}\over \tau^2} \quad \Rightarrow \quad
\nu^2 = {9 \over 4} + 6 \; \epsilon - 3 \beta - 9 \epsilon \; \beta + \beta^2 + 2 \epsilon \; \beta^2
\end{equation}
In general, the solutions to \eqref{Bessel} are
\begin{eqnarray}
f_k = c_1 (k) H^{(1)}_{\nu} (-k \tau) + c_2 (k) H^{(2)}_{\nu} (-k \tau)
\end{eqnarray}
Here $ H^{(1)}_{\nu}$ and  $ H^{(2)}_{\nu}$ are   the  Hankel's  functions of the first and second kind, respectively, and $c_1 (k)$ and $c_2 (k)$ are arbitrary constants. Comparing the asymptotic behavior of the general solution, with the solution of the equation in the sub-horizon limit ($k\tau \ll 1$), the constants are determined, and finally one could obtain the amplitude of the scalar perturbations as
\begin{equation}\label{SAmp}
\mathcal{P}_s = A^2_s \; \left( k \over a H \right)^{3-2\nu}, \qquad
A_s^2 = {1 \over 25\pi^2} \left( 2^{\nu-3/2} \Gamma(\nu) \over \Gamma(3/2) \right)^2 \; \left( {H^2 \over \dot\phi} \right)^2.
\end{equation}
from which we deduce    the scalar spectral index $n_s$  as
\begin{equation}\label{ns}
n_s-1 = 3 -2\nu
\end{equation}

\subsection{Tensor perturbations}
The second SRP plays no role in the tensor perturbation equations, and hence  the evolution equation for  the  tensor perturbation will have the same form as the scalar case.  The amplitude of such perturbations has been calculated in  the framework of the brane-world gravity and is given by \cite{Langlois:2000ns,Huey:2001ae}
\begin{equation}\label{TAmp}
A_T^2 = {16 \pi \over 25 M_p^2} \; \left( {H \over 2\pi} \right)^2 \; F^2(x),
\end{equation}
where
\begin{equation}
F^2 = \left[ \sqrt{1+x^2} \; - x^2 \sinh^{-1}\left( {1 \over x} \right) \right]^{-1}, \qquad
 x \equiv H M_p \sqrt{3 \over 4\pi \lambda}.
\end{equation}
In high energy regime, where $x \gg 1$, one arrives at $F(x)= 3x/2$ \cite{Huey:2001ae,Videla:2016nlt}. The tensor perturbations are measured indirectly through the parameter $r$, defined as the ratio of tensor perturbations to scalar perturbations, which can be determined  using Eqs.\eqref{SAmp} and \eqref{TAmp} as
\begin{equation}\label{r}
  r = {3 \over 2} \; \left( \Gamma(3/2) \over 2^{\nu-3/2} \Gamma(\nu) \right)^2 \; \epsilon.
\end{equation}
Currently, the value of this parameter  is not determined by  the data,  and only   an upper bound  $r < 0.064$ \cite{Akrami:2019izv,Akrami:2018odb,Array:2015xqh}.

\section{Observational constraints on the model}
\begin{figure}
  \centering
  \includegraphics[width=7cm]{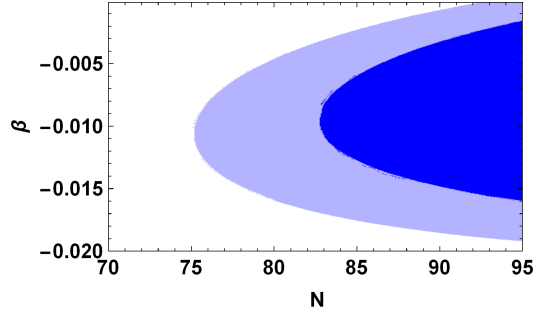}
  \caption{The $68\%$ (light blue) and $95\%$ CL (dark blue) allowed  region of the  parameters   $\beta$ and $N$.}\label{betaN}
\end{figure}

To determine the  free  parameters of the model, we compute the amplitude of the scalar perturbations, scalar spectral index, and tensor-to-scalar ratio at the time of horizon crossing  and compare with the available observational data. First, by  substituting   the expression in Eq.\eqref{phistar} into  Eq.\eqref{epsilon},   the slow-roll parameter $\epsilon$ can be written  in terms of the number of e-folds as
\begin{equation}\label{epsilonS}
  \epsilon = {-2 \beta \; e^{2\beta N} \over 1 - 2\beta - e^{2\beta N}}.
\end{equation}
Note  that  (from Eqs.\eqref{nu}, \eqref{ns}, and \eqref{r}) the scalar spectral index and tensor-to-scalar ratio depend only on $\beta$ and $N$ at the time of horizon crossing. Comparing the theoretical results for $n_s$ and $r$ with allowed values  of the spectral index and tensor-to-scalar ratio given  by Planck collaboration  in the form of  $r-n_s$ diagram, we  extract the  values of  $(N,\beta)$ that  are  in agreement with  the observational data.  Using the $95\%$  and $68\%$ CL  allowed regions  of the parameters $r$ and $n_s$  from  Planck TT,TE,EE+lowE+lensing+BK14+BAO datasets\cite{Akrami:2018odb}, we show in Fig.\ref{betaN} the  corresponding model parameter space.

Using the amplitude of the scalar perturbations, the other constant of the model, i.e. $\alpha$, is determined as
\begin{equation}\label{alpha}
  \alpha^3 = \left( \Gamma(3/2) \over 2^{\nu-3/2} \Gamma(\nu) \right)^2 \; \left( {4\pi \over M_5^3} \right)^2 \;
   \Big( 4\pi \; (2\beta)^3 \; A_s \; \epsilon \Big)
\end{equation}
To have numerical insight about the result of the model, Table.\ref{T01} represents the values of $\alpha$, scalar spectral index, tensor-to-scalar ratio, and the energy scale of inflation for different values of $\beta$ and  the number of e-fold, taken from Fig.\ref{betaN}.
\begin{table}
  \centering
  \begin{tabular}{p{1.7cm}p{1cm}p{2.5cm}p{1.5cm}p{1.5cm}p{2cm}}
    \hline
    \quad \ $\beta$ & $N$ & \qquad $\alpha$ & \quad $n_s$ & \quad $r$ & \qquad $V^\star$ \\
    \hline
    $-0.011$ & $76$ & $4.92 \times 10^{-33}$ & $0.9580$ & $0.0072$ & $2.22 \times 10^{53}$ \\

    $-0.014$ & $80$ & $4.95 \times 10^{-33}$ & $0.9589$ & $0.0047$ & $1.92 \times 10^{53}$ \\

    $-0.007$ & $80$ & $4.16 \times 10^{-33}$ & $0.9594$ & $0.0096$ & $2,45 \times 10^{53}$ \\

    $-0.014$ & $84$ & $4.64 \times 10^{-33}$ & $0.9604$ & $0.0041$ & $1.85 \times 10^{53}$ \\

    $-0.010$ & $84$ & $4.32 \times 10^{-33}$ & $0.9620$ & $0.0065$ & $2.15 \times 10^{53}$ \\

    $-0.004$ & $84$ & $3.51 \times 10^{-33}$ & $0.9592$ & $0.0119$ & $2.62 \times 10^{53}$ \\

    $-0.009$ & $88$ & $4.01 \times 10^{-33}$ & $0.9637$ & $0.0066$ & $2.16 \times 10^{53}$ \\
    \hline
  \end{tabular}
  \caption{numerical results of the model}\label{T01}
\end{table}

\begin{figure}[h]
  \centering
  \includegraphics[width=7cm]{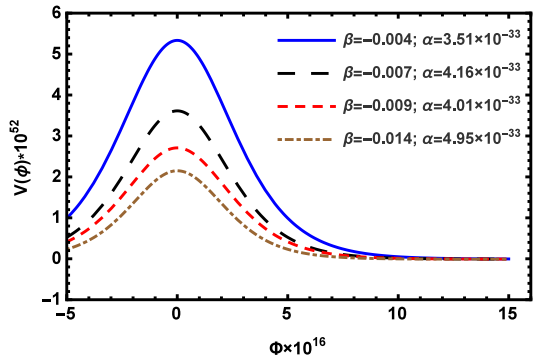}
  \caption{Behavior of the potential of the scalar field.}\label{potential}
\end{figure}

Fig.\ref{potential} portrays the behavior of the obtained potential versus the scalar field for different values of $\beta$ and $\alpha$. As it is illustrated, the potential rolls down from the top of the potential.

The crucial point for any inflationary model is to have a graceful exit from the inflation stage. Considering the behavior of the first SRP presents the required information about the inflationary times and its end. The evolution of $\epsilon$ versus the number of e-fold  is depicted in Fig.\ref{epsilon}, where it is realized that by approaching the end of inflation the parameter $\epsilon$ grows up and reaches one.
\begin{figure}[h]
  \centering
  \includegraphics[width=7cm]{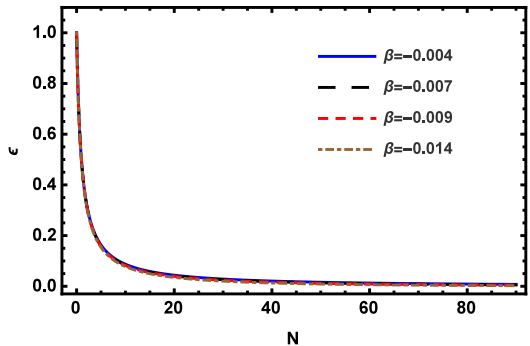}
  \caption{Behavior of the first slow-roll parameter $\epsilon$ versus the number of e-fold.}\label{epsilon}
\end{figure}

\section{Consistency with the Swampland Criteria}
\begin{figure}[h]
  \centering
  \subfigure[]{\includegraphics[width=7cm]{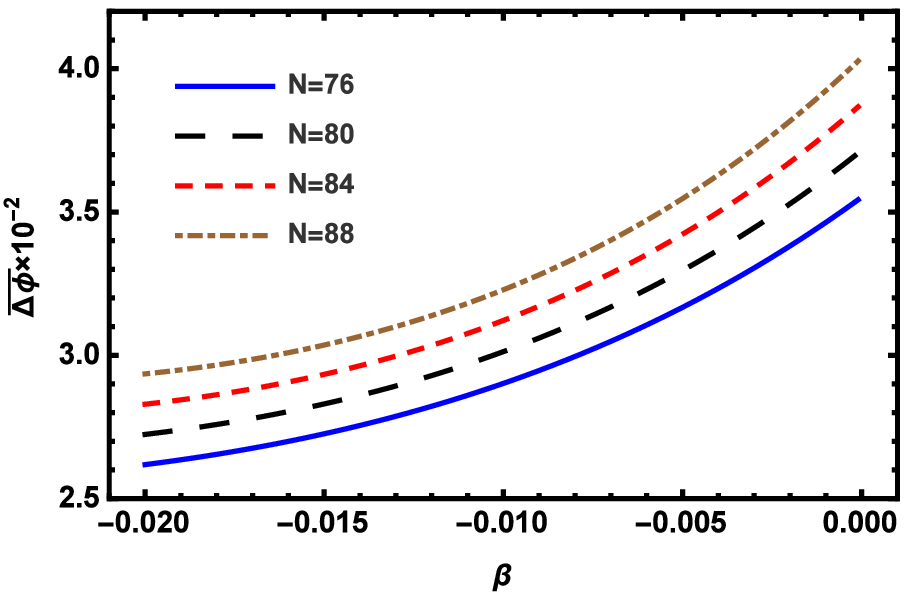}}
  \subfigure[]{\includegraphics[width=6.5cm]{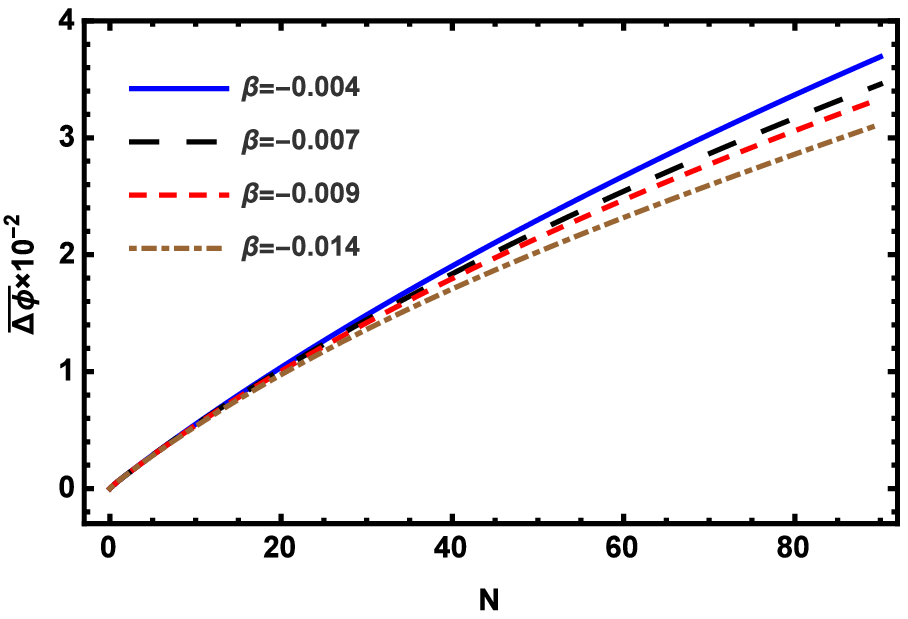}}
  \caption{Evolution of $\Delta \phi/M_p$ versus the number of e-fold for different values of $\beta$.}\label{deltaphi}
\end{figure}

\begin{figure}[h]
  \centering
  \subfigure[]{\includegraphics[width=7cm]{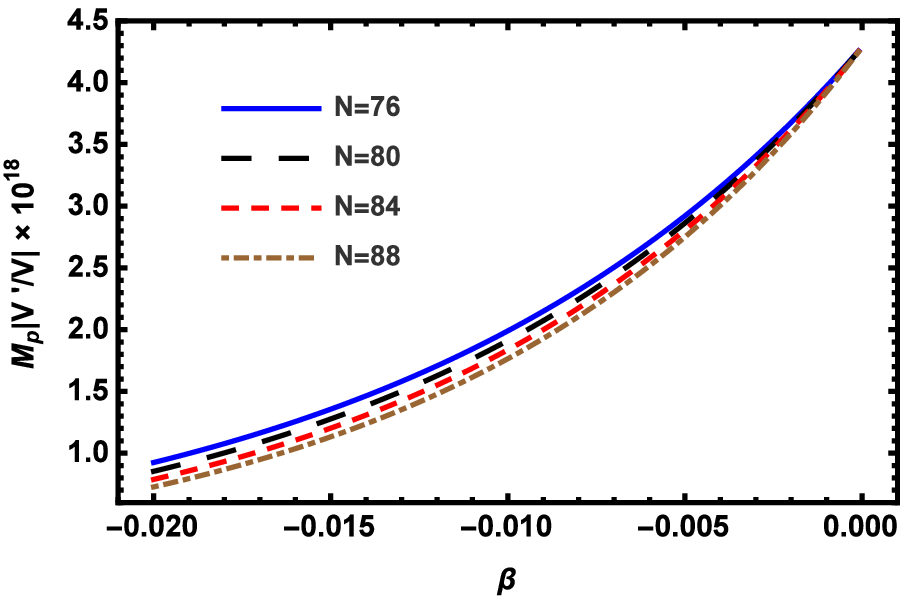}}
  \subfigure[]{\includegraphics[width=6.5cm]{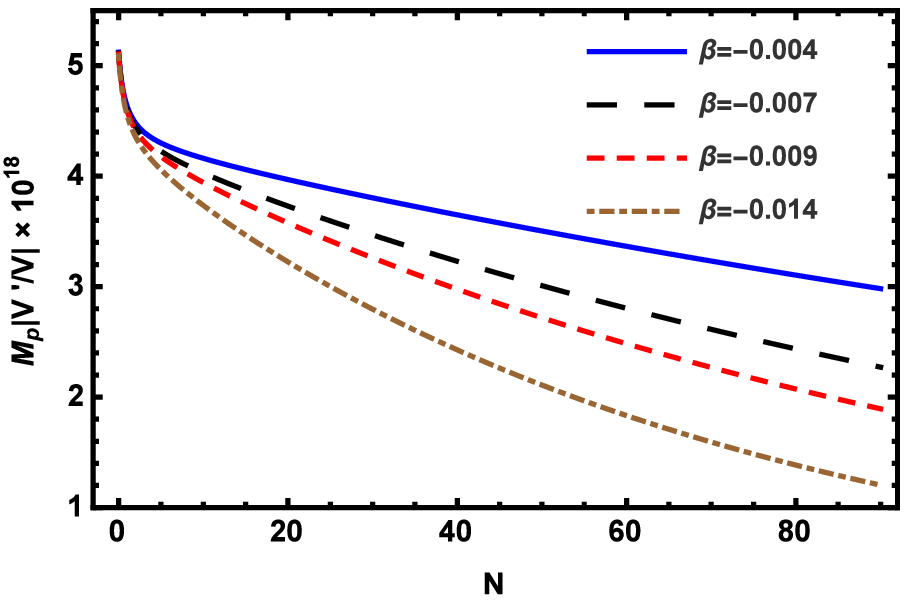}}
  \caption{Evolution of the gradient of the potential versus the number of e-fold for different values of $\beta$.}\label{dV}
\end{figure}

The recently proposed swampland criteria is a measure for separating the consistent EFT from the inconsistent EFT. The consistent EFTs can successfully be formulated in  string theory,  the best candidates of the quantum gravity. It is believed Inflation occurred at the energy scale below the Planck energy and hence   could be described by a low-energy effective field theory  of string theory. Therefore, it is a  natural desire to construct an inflationary model based on a consistent EFT, and for that we apply  the swampland conjectures.\\
The first criterion concerns the distance conjecture which constraints on  the range traversed by the scalar field as  $\Delta \phi / M_p < c$ where $c$ is of the order  of unity. The evolution of the term $\Delta \phi / M_p$ for the model is presented in Fig.\ref{deltaphi}, where it is shown  that $\Delta \phi / M_p$ is smaller than unity for the whole time of the inflation. \\
The second criterion is a de Sitter conjecture which imposes a lower bound on the gradient of the potential. It states that $M_p |V'/V|>c'$ where $c'$ is of the order of unity (further investigation determines that the constant could be of the order of $0.1$ \cite{Kehagias:2018uem}. In Fig.\ref{dV}, we present the evolution  of  $M_p |V'/V|$, which shows that the magnitude of the gradient of the potential is bigger than one during the inflationary phase. \\

\section{Attractor behavior}
\begin{figure}[h]
  \centering
  \includegraphics[width=7cm]{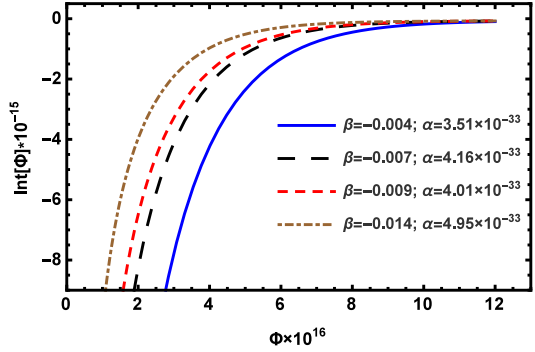}
  \caption{The curves display the behavior of the integrand versus the scalar field during the inflationary time for different values of $\beta$ and $\alpha$.}\label{attractor}
\end{figure}

The last feature we are going to consider is the attractor behavior of the model. The solution of the model has been obtained in Sec.III, where we have used the Hamilton-Jacobi formalism \cite{Salopek:1992qy,Liddle:1994dx,Lyth:2009zz,Liddle:2000cg,Kinney:1997ne,Guo:2003zf,Aghamohammadi:2014aca,Saaidi:2015kaa,Sheikhahmadi:2016wyz,Achucarro:2018ngj}.
 This approach  was first studied in \cite{Motohashi:2014ppa}, where the authors found  exact solution for the large value of the parameter $\eta$. Considering the attractor behavior of the solution, it was claimed that constant-roll inflation presents a new class of attractor solution. This result has been re-investigated in detail in  \cite{Morse:2018kda,Lin:2019fcz}  where it was  shown  that the solution and the perturbation equations are invariant under the  transformation $\eta \rightarrow \bar{\eta} = 3-\eta$, with two branches of solution  that are symmetric under this transformation. The main result of \cite{Morse:2018kda} based on this duality transformation led the authors to the conclusion that the attractor behavior of the constant-roll inflation with large $\eta$ is not a new class of attractor behavior which also has been  claimed in \cite{Morse:2018kda}.  It  does not mean  that the constant-roll has no attractor solution, but   they are not a new class of attractor solution. \\
In our model we found  that the observational constraints on $(r, n_s)$ requires the  parameter $\eta(=\beta)$ to be small. Consequently, 
 the model certainly does not present a new class of the attractor solution, and    could be part of the slow-roll attractor. Therefore, the attractor behavior in our model could be investigated utilizing the same method as  in the slow-roll scenario. In this regard, we follow a similar  procedure as in  \cite{Lyth:2009zz,Liddle:1994dx} which is a common method for considering the attractor behavior of the inflationary models.    \\
Assuming  homogenous perturbation in the Hubble parameter, i.e. $H(\phi) = H_0 + \delta H(\phi)$, and substituting it into  the Hamilton-Jacobi equation,
\begin{equation}\label{HJequation}
  V(\phi) = \left( 3M_5^3 \over 4\pi \right) \; H(\phi) - {1 \over 9} \; \left( 3M_5^3 \over 4\pi \right)^2 \; {H'^2(\phi) \over H^2(\phi)},
\end{equation}
leads to the following differential equation
\begin{equation}\label{deltaH}
  {\delta H' (\phi) \over \delta H (\phi)} =  \left( 1 + {9 \over 2} \; \left( 4\pi \over 3M_5^3 \right) \; {H_0^2 \over H^{\prime 2}_0} \right) \; {H'_0 \over H_0}
\end{equation}
where the equation has been obtained up to the first order of the perturbation term. Integration lead to
\begin{equation}\label{deltaHphi}
  \delta H(\phi) = \delta H_i \; \exp\left[ \int_{\phi_i}^{\phi} \left( 1 + {9 \over 2} \;\left( 4\pi \over 3M_5^3 \right) {H^3_0(\phi) \over H'^2_0(\phi)}  \right) \; {H'_0(\phi) \over H_0(\phi)} \; d\phi \right]
\end{equation}
The integrand is illustrated in Fig.\ref{attractor} versus the scalar field. The curves portray the behavior of the integrand versus the scalar field during the inflationary times. The area between the curve and the x-axis displays the actual value of the integral in the power of the exponential term in Eq.\eqref{deltaHphi}. Inflation begins for smaller field and it ends at bigger fields. Therefore, as the time  passes and approaches the end of inflation, the area under the curve is getting larger  and larger and the integral becomes more and more negative. Then, the exponential term approaches to zero implying that the homogeneous perturbation $\delta H(\phi)$ dies away with time, and the model possesses attractor behavior.

\section{Conclusion}
The constant-roll inflation was investigated in the frame of the RSII brane gravity model. Based on this scenario, the universe and all the matter fields, including the inflaton,   are confined to a brane with positive tension,  where  the brane is embedded in five-dimensional space-time. The modified gravity model results in a modified Friedmann equation which contains  both linear and quadratic  terms of the energy density. In the high energy limit, the quadratic term dominates, and consequently, the Hubble parameter becomes  proportional to the energy density $\rho$, instead of $\sqrt{\rho}$.  In this scenario, the inflaton rolls down its potential at a constant rate  where   the second slow-roll inflation parameter  is taken to be  constant which, in general,  can  be of order unity. Using the Hamilton-Jacobi approach,  we derive a differential equation for the Hubble parameter. For our model, there is a non-linear second-order differential equation that gives an exact solution for the model. Finding the Hubble parameter in terms of the scalar field, the other background parameters, such as the time derivative of the scalar field and the potential, were derived in terms of the scalar field.  The slow-roll parameter $\epsilon$ was also obtained in terms of the scalar field, which is used to infer  the scalar field at the end of inflation through the relation $\epsilon(\phi_e) = 1$. The scalar field at the beginning of inflation was acquired from the expression  of the number of e-fold.  \\
Another consequence of this scenario appears in the perturbation parameters where one could find the modified terms mainly in the amplitude of the scalar perturbations, scalar spectral index, and tensor-to-scalar ratio. Since the second slow-roll parameter might not be small, the scalar perturbation equations were reconsidered, and the modified scalar power spectrum was derived. The tensor power spectrum is the  same as the slow-roll in brane inflation because the second slow-roll inflation parameters play no role in tensor perturbation equations. \\
Computing the perturbation parameters at the time of horizon crossing, the scalar spectral index and tensor-to-scalar ratio are obtained only in terms of the constant $\beta$ (i.e. the second slow-roll parameter) and the number of e-fold. Comparing the theoretical results of the model with the Planck data, a set of the $(\beta,N)$ is found that four any point in this set, the model perfectly agrees with observational data. The other constant of the model, i.e. $\alpha$, is determined from the amplitude of the scalar perturbation where there is an exact value for the parameter based on data. Using this result, a numerical result of the model about the main parameters including the energy scale of inflation are presented. In the next step, the consistency of the model with the recently proposed swampland criteria is considered. We tried to find whether the model with the obtained free parameter could satisfy the conjectures. Furthermore,   the range of the scalar field values  and the gradient of its potential  appropriately satisfy both swampland criteria. \\
Finally, in the Hamilton-Jacobi formalism, we derived the differential equation (up to the first order) describing the behavior of a homogeneous perturbation for the Hubble parameter as a function of the inflaton field. We showed   that the perturbation parameter reduces as the time approaches the end of inflation, which indicates that the solution of the model has the attractor behavior. \\


\acknowledgments
We would like to thank  H. Azri for careful  reading of the manuscript. The work of A.M. has been supported financially by Vice Chancellorship of Research and Technology, University of Kurdistan under research Project No.98/10/34704. The work of T. G. has been supported financially by Vice Chancellorship of Research and Technology, University of Kurdistan‘under research Project No.98/11/2724.

\bibliography{BCRref}

\end{document}